\documentclass{aa}

\usepackage{amssymb}
\usepackage{amsmath}
\usepackage{txfonts}
\usepackage{graphicx}
\usepackage{xspace}
\usepackage{natbib}

\newcommand{\fu}{4U\,1907+09}
\newcommand{\inte}{\textsl{INTEGRAL}}
\newcommand{\xte}{\textsl{RXTE}}
\newcommand{\sax}{\textsl{BeppoSAX}}
\newcommand{\ginga}{\textsl{Ginga}}
\newcommand{\kev}{\ensuremath{\text{keV}}}
\newcommand{\asm}{\textsl{ASM}}
\newcommand{\msun}{\ensuremath{\text{M}_{\odot}}}
\newcommand{\rsun}{\ensuremath{\text{R}_{\odot}}}
\newcommand{\lsun}{\ensuremath{\text{L}_{\odot}}}
\newcommand{\redchi}{\ensuremath{\chi^{2}_\text{red}}}

\begin{document}
\title{A torque reversal of \fu}

\author{S.~Fritz\inst{1} \and I.~Kreykenbohm\inst{1,2} \and
  J.~Wilms\inst{3}\fnmsep\thanks{\emph{present address:} Dr.\ Remeis
  Sternwarte, Astronomisches Institut der Universit\"at
  Erlangen-N\"urnberg, Sternwartstr.~7, 96049 Bamberg, Germany}
  \and R.~Staubert\inst{1} \and F.~Bayazit\inst{1} \and 
    K.~Pottschmidt\inst{4} \and J.~Rodriguez\inst{5} \and
    A.~Santangelo\inst{1}} 

\offprints{S. Fritz, \\ \email{fritz@astro.uni-tuebingen.de}}

\institute{ Institut f\"ur Astronomie und Astrophysik, Sand 1, 72076
 T\"ubingen, Germany \and  
 INTEGRAL Science Data Centre, 16 Ch.\ d'\'Ecogia, 1290 Versoix,
 Switzerland \and  
 Department of Physics, University of Warwick, Coventry, CV4 7AL,
 U.K. \and
 CASS, University of California at San Diego, La Jolla, CA 92093-0424, USA \and
CEA Saclay, Unit\'e Mixte de Recherche  CEA-CNRS-Universit\'e Paris 7  
(UMR 7158/AIM) - DSM/DAPNIA/SAp,
F-91191 Gif sur Yvette, France}

\date{Received 9 May 2006 ; Accepted 8 August 2006 } 

\titlerunning{Torque reversal of \fu}
 
\abstract{}{We present an analysis of the accreting X-ray pulsar system \fu\  
  based on \inte\ data. The main focus of this analysis is a study of the
  timing behavior of this source. In addition we also show an analysis of the
  5--90\,\kev\ spectrum.}  {The data were extracted using the official \inte\
  software OSA~5.1. Timing analysis was performed using epoch folding and
  pulsar pulse phasing.}  {We have measured 12 individual pulse periods for
  the years 2003 to 2005. We confirm earlier \xte\ results that during 2003 the
  spin down became slower and show furthermore that after this phase \fu\ 
  started to spin up with
  $\dot{P}_\text{pulse}=-0.158\,\text{s}\,\text{yr}^{-1}$ in 2004. The
  similarity of the pulse period histories of \fu\  and 4U\,1626$-$26 suggests
  that accretion onto an oblique rotator, as recently proposed by Perna et
  al., is a possible explanation for this change.}{}
  

\maketitle
\section{Introduction}\label{sect:intro}
\begin{figure}  
  \resizebox{\hsize}{!}{\includegraphics{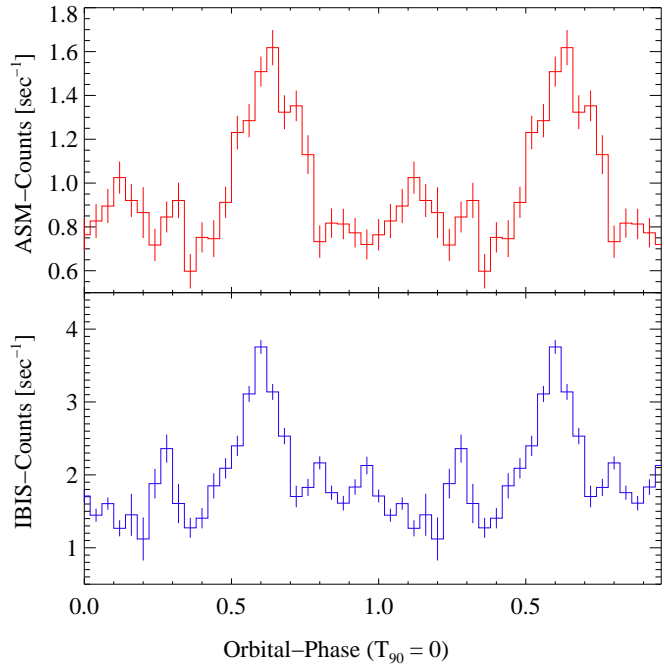}} 
  \caption{Upper Panel: Rossi X-ray Timing Explorer All Sky Monitor
    (ASM) 2--12\,keV light curve of \fu\ folded on the orbital period
    of 8.3753\,days. Lower Panel: Same for the IBIS 20--40\,keV light
    curve.}\label{fig:asm}
\end{figure}

The wind-accreting High Mass X-ray Binary system \object{\fu}
\citep{giacconi71} consists of a neutron star in an eccentric
($e=0.28$) 8.3753\,d orbit \citep{intzand98} around its companion,
which has been identified optically with a highly reddened
$m_\text{V}=16.37$\,mag star \citep{schwartz80}.  Using interstellar
atomic lines of \ion{Na}{i} and \ion{K}{i}, \citet{cox05} set a lower
limit of 5\,kpc for the distance. According to this value a lower
  limit of the X-ray luminosity above 1\,\kev\ is given by 
$2\,10^{36}\,\text{erg}\,\text{s}^{-1}$ \citep{intzand97}. 
\citet{cox05} also confirm earlier suggestions
\citep{schwartz80,marshall80,kerkwijk89} that the stellar companion is a
O8--O9 Ia supergiant with an effective temperature of 30500\,K, a radius of
26\,\rsun, a luminosity of $5\,10^{5}\,\lsun$, and a mass loss rate of
$7\,10^{-6}\,\msun\,\text{yr}^{-1}$. Note that the presence of X-ray flaring
seen twice per neutron star orbit \citep[][ see also
Fig.~\ref{fig:asm}]{marshall80} had led some authors
\citep[e.g.,][]{makishima84,iye86,cook87} to the suggestion of a Be star
companion, however, this classification would require a distance of
$<$1.5\,kpc, which is in contradiction to the significant interstellar
extinction measured in optical observations \citep{kerkwijk89}.

Similar to other accreting neutron stars, the X-ray continuum of \fu\  can be
described by a power-law spectrum with an exponential turnover at
$\sim$13\,keV \citep{mihara95,intzand97,cusumano98}.  The spectrum is modified
by strong photoelectric absorption with a column $N_\text{H}=1.5$--$5.7\,
10^{22}\,\text{cm}^{-2}$
\citep{schwartz80,marshall80,makishima84,cook87,chitnis93,cusumano98}.  As the
absorbing material is situated in the dense stellar wind, $N_\text{H}$ is
strongly variable over the orbit.  The column is maximal between the end of
the primary X-ray flare and the start of the secondary flare
(Fig.~\ref{fig:asm}) of the orbital light curve
\citep{roberts01}. Fluorescence in the absorbing material gives rise to an Fe
K$\alpha$ line at 6.4\,keV with an equivalent width of $\sim$60\,eV
\citep{cusumano98}.  At higher energies, the spectrum exhibits cyclotron
resonant scattering features (CRSF) at $\sim$19\,\kev\ and $\sim$40\,keV
\citep{makishima92,mihara95,cusumano98}.

With a pulse period of $\sim440$\,s, \fu\  is a slowly rotating neutron
star.  Since the discovery of the pulsations by \citet{makishima84}
the neutron star has exhibited a steady linear spin down with an
average of $\dot{P}_\text{pulse}=+0.225\,\text{s}\,\text{yr}^{-1}$
from $P_{\text{pulse}}=437.5$\,s in 1983 to 440.76\,s in 1998
\citep{intzand98}. Recently, \citet{baykal05} reported a decrease in
$\dot{P}_\text{pulse}$, which in 2002 was
$\sim$$0.115\,\text{s}\,\text{yr}^{-1}$ and therefore $\sim$0.5 times
the long term value.

In this paper we report on an analysis of \inte\ observations of \fu\ 
performed since 2003. In Sect.~\ref{sec:datareduction} we describe the
\inte\ data reduction, followed by a discussion of the broad-band X-ray
spectrum of the source (Sect.~\ref{sec:specevol}).  We then present
the results of the long term X-ray timing analysis
(Sect.~\ref{sec:lc}), showing that the source underwent a torque reversal
towards a spin up (Sect.~\ref{sec:pulsevol}). We summarize and discuss
our results in Sect.~\ref{sec:summary}.

\section{Observation and data reduction}\label{sec:datareduction}
In this paper we present the analysis of \inte\ observations of \fu.  The
International Gamma-Ray Astrophysics Laboratory (\inte; \citealt{winkler03})
was launched in October 2002. The two main instruments IBIS
(15\,\kev--10\,MeV; \citealt{ubertini03}) and SPI (20\,\kev--8\,MeV;
\citealt{vedrenne03}) are supplemented in the X-rays by an auxiliary
instrument with a smaller field of view, the X-ray monitor JEM-X
(3\,\kev--35\,\kev; \citealt{lund03}).  For our analysis we used data
from IBIS and JEM-X.

The Imager on Board the \inte\ Satellite (IBIS) is a coded mask telescope with
a fully coded field of view (FCFOV) of $9^\circ\times9^\circ$ and
$12\farcm{}9$ full width at half maximum (FWHM) angular resolution
\citep{ubertini03,brandt03}.  Here, we used data from its upper layer, the
\inte\ Soft Gamma-Ray Imager (ISGRI), which covers an energy range from
15\,\kev--1\,MeV with an energy resolution of $\sim$8\% at 60\,\kev\
\citep{lebrun03,gros03}. \fu\  is detected below $\sim$90\,keV. The Joint
European X-ray Monitor (JEM-X) consists of two identical coded mask
instruments with a FCFOV of $4\fdg8$ diameter and $3\farcm{}35$ angular
resolution (FWHM), each \citep{lund03}. Due to erosion of the microstrip
anodes inside the JEM-X detectors which leads to a loss in sensitivity, only
one of the two JEM-X detectors is operating at any given time while the other
is switched off.

For our analysis we took into account all public data up to \inte\
orbit (revolution) 250 as well as data from the \inte\ Galactic Plane
Scans (GPS) and the Galactic Central Radian Deep Exposure (GCDE). In
addition we also used private data from a monitoring campaign on the
Galactic microquasar GRS~1915+105 (PI J.~Rodriguez). To avoid
systematic effects that start to become important for larger off-axis
angles, we constrained the maximum off-axis angle between the
satellite's optical axis and \fu\  to $4\fdg5$ for IBIS and $2\fdg4$ for
JEM-X according to the respective FCFOVs of the instruments.
In total, the analyzed data are spread over almost three years with a
total on source time of $\sim$2280\,ksec for IBIS.  The
log of the observations is given in Table~\ref{tab:obs}.  We
used the Off-line Scientific Analysis Software, \texttt{OSA}\,5.1, in
our analysis. In the analysis of the coded mask data we also took into
account the presence of \object{GRS~1915+105}, \object{4U\,1901+03},
\object{4U\,1909+07}, and \object{IGR~J19140+0951}, by forcing the
\texttt{OSA} to consider \fu\  and these four sources.

\begin{figure}  
  \resizebox{\hsize}{!}{\includegraphics{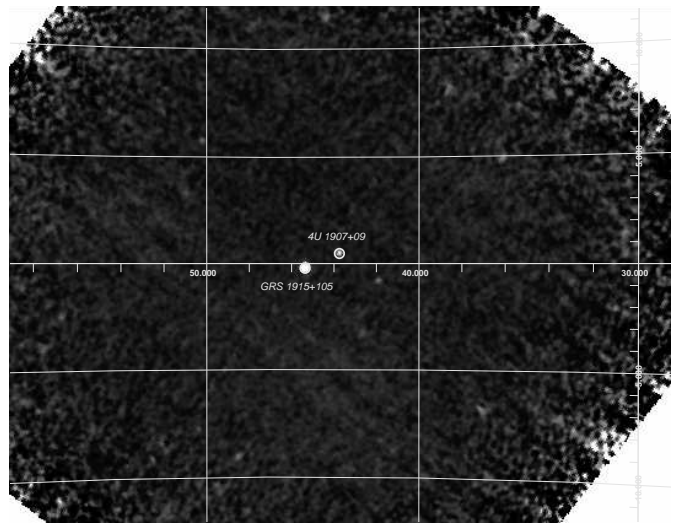}}
  \caption{ISGRI image of the \fu\ region in the 20--40\,keV band (exposure
  time about 100\,ksec, data from revolution 246).}
   \label{fig:image}
\end{figure}

Background subtracted light curves were obtained using \texttt{ii\_light}, a
  tool distributed with the \texttt{OSA} to generate high resolution light
  curves. While the standard \texttt{OSA} tools build lightcurves by
  deconvolving shadowgrams for each requested energy band and time bin to
  obtain the light curves, which is only possible if the signal to noise ratio
  in the shadowgram is high, \texttt{ii\_light} uses the Pixel Illuminated
  Fraction (PIF) to create light curves.  For \fu, this approach allows us to
  extract light curves with a resolution of 1\,s.  The count rates found by
  \texttt{ii\_light} show no systematic difference between the single
  pointings. Tests based on observations of the Crab nebula and pulsar
  \citep{kreykenbohm06} and the fact that the count rates of \fu\ are not
  affected by the very variable flux of the nearby source GRS 1915+105
  (Fig.~\ref{fig:image}) confirm the stability of our extraction method.  The
  light curves were barycentered and corrected for the orbital motion of the
  neutron star using the ephemeris of \citet{intzand98}, as given in
  Table~\ref{tab:eph}.  Orbital phase 0 is at $T_{90}$.

\begin{table}
\caption{Ephemeris used for the binary correction
    \citep{intzand98}. $P_\text{orb}$ is the orbital period, $T_{90}$ 
    the time of mean longitude $90^\circ$. All other symbols have their usual meanings.}
\label{tab:eph}
\centering
\begin{tabular}{ll}
\hline\hline
Parameter          & Value         \\
\hline
    $T_{90}$       & MJD\,50134.76 \\ 
    $P_\text{orb}$ & 8.3753\,d     \\ 
    $a\sin{i}$     & 83\,lt-s      \\ 
    Eccentricity   & 0.28          \\ 
    $\omega$       & $330^\circ$   \\ 
\hline
\end{tabular}
\end{table}

\begin{table*}
\begin{minipage}{\linewidth}
\centering
\caption{Log of observations.}  
\label{tab:obs}
\begin{tabular}{cccccc}
\hline\hline
Observation & Revolution & Date & On Source Time (IBIS) & $\Phi_\text{orb}$ & Mean
Count Rate \\
  &                & MJD              & ksec   &  & cps in 20--40\,keV \\
\hline
\phantom{0}1 & \phantom{0}48  & 52704.1--52705.4 & 100.61 & 0.78--0.93  & 1.26$\pm$0.10\\
\phantom{0}2 & \phantom{0}57  & 52731.1--52732.3 & \phantom{0}72.12  & 0.00--0.14  & 2.44$\pm$0.13\\
\phantom{0}3 & \phantom{0}59  & 52738.3--52739.6 & 107.13 & 0.86--0.02  & 1.66$\pm$0.11\\
\phantom{0}4 & \phantom{0}62  & 52746.7--52747.9 & \phantom{0}72.78  & 0.86--0.01  & 2.76$\pm$0.13\\
\phantom{0}5 & \phantom{0}67  & 52762.9--52763.4 & \phantom{0}33.48  & 0.80--0.86 & 2.12$\pm$0.19\\
\phantom{0}6 & \phantom{0}68  & 52764.6--52766.5 & \phantom{0}61.66  & 0.00--0.22 & 1.92$\pm$0.14\\
\phantom{0}7 & \phantom{0}69  & 52767.8--52769.5 & 107.41 & 0.38--0.59 & 1.66$\pm$0.11 \\
\phantom{0}8 & \phantom{0}70  & 52771.2--52771.7 & \phantom{0}32.71  & 0.79--0.85  & 1.87$\pm$0.19\\
\phantom{0}9 & 135 & 52965.7--52966.9 & 104.37 & 0.01--0.16  & 1.50$\pm$0.12\\
10& 172 & 53075.9--53076.2 & \phantom{0}23.03  & 0.17--0.21 & 1.39$\pm$0.23\\
11& 174 & 53082.6--53083.0 & \phantom{0}23.09  & 0.97--0.01 & 2.26$\pm$0.24\\
12& 176 & 53089.1--53089.6 & \phantom{0}30.83  & 0.74--0.81 & 1.07$\pm$0.20\\
13& 177 & 53091.0--53092.0 & \phantom{0}51.80  & 0.97--0.09 & 1.30$\pm$0.15\\
14& 186 & 53117.0--53117.7 & \phantom{0}37.18  & 0.07--0.16 & 2.37$\pm$0.18\\
15& 187 & 53120.7--53122.0 & \phantom{0}65.47  & 0.52--0.67 & 3.70$\pm$0.14\\
16& 188 & 53123.2--53125.4 & \phantom{0}75.72  & 0.82--0.08 & 2.55$\pm$0.13\\
17& 189 & 53126.3--53128.5 & \phantom{0}62.89  & 0.19--0.45 & 1.46$\pm$0.14\\
18& 193 & 53138.1--53140.5 & 185.49 & 0.60--0.88 & 3.60$\pm$0.09\\
19& 231 & 53253.4--53254.0 & \phantom{0}33.69  & 0.36--0.43 & 1.32$\pm$0.19\\
20& 242 & 53285.4--53285.6 & \phantom{0}19.16  & 0.18--0.21 & 2.02$\pm$0.26\\
21& 243 & 53288.9--53289.2 & \phantom{0}21.24  & 0.60--0.64 & 2.27$\pm$0.24\\
22& 246 & 53296.4--53297.6 & 100.89  & 0.49--0.64 & 2.79$\pm$0.11\\
23& 248 & 53303.9--53305.0 & 144.48  & 0.27--0.52 & 1.78$\pm$0.09\\
24& 249 & 53305.7--53308.0 & 141.95  & 0.61--0.87 & 1.54$\pm$0.09\\
25& 250 & 53309.1--53311.0 & 100.83  & 0.02--0.23 & 0.97$\pm$0.12\\
26& 255 & 53324.3--53325.5 & \phantom{0}98.74  & 0.82--0.97 & 1.60$\pm$0.11\\
27& 295 & 53443.0--53444.1 & \phantom{0}99.40  & 0.99--0.14 & 1.37$\pm$0.11\\
28& 305 & 53472.9--53474.1 & 106.13 & 0.57--0.72 & 3.11$\pm$0.11\\
29& 315 & 53503.6--53504.8 & \phantom{0}72.15 & 0.24--0.37 & 1.88$\pm$0.17\\
30& 361 & 53640.4--53641.6 & \phantom{0}91.67 & 0.57--0.71 & 3.16$\pm$0.11\\
\hline
\end{tabular}
\end{minipage}
\end{table*}

\section{Spectral Analysis}\label{sec:specevol}
\begin{figure}  
  \resizebox{\hsize}{!}{\includegraphics{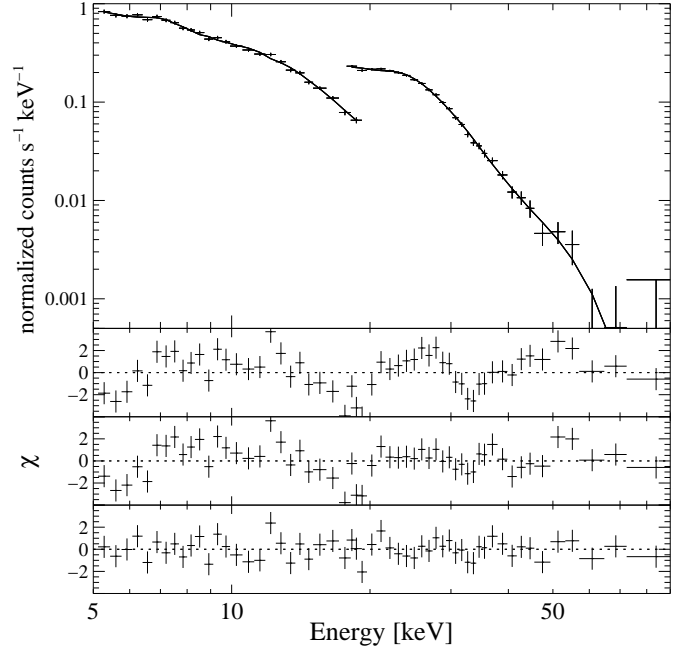}}
  \caption{Fit to the data using the Fermi-Dirac cutoff model
    including 2 cyclotron lines. The bottom panels show the residuals,
    from top to bottom the case without lines, with 1 line, and with 2
    cyclotron lines.}
   \label{fig:spe}
\end{figure}

For the spectral analysis we applied systematic errors of 1\% and 5\%
for IBIS and JEM-X2, respectively, and used data taken in the energy
ranges 18--90\,\kev\ and 5--20\,\kev.  Because of the switch over from
JEM-X2 to JEM-X1 after revolution 171, we only used data taken before
that revolution to avoid cross-calibration uncertainties
  between the two JEM-X detectors. This resulted in a total exposure
time of 200\,ksec for JEM-X2. For ISGRI we used all public data up to
revolution 250 resulting in an exposure time of 1258\,ksec. We modeled
the spectrum using an absorbed power-law, which is modified by the
Fermi-Dirac cutoff \citep{tanaka86}:
\begin{equation}
 C(E)=E^{-\alpha}\,\left[\exp\left(\frac{E_\text{cut}-E}{E_\text{fold}}\right)+1\right]^{-1}
\end{equation}
This continuum model alone could not describe the data adequately
  ($\chi_\text{red}^{2}=3.10$, see Tab.~\ref{tab:fitfdcut}), as strong
  absorption line like features at $\sim$19\,keV and $\sim$40\,keV remain
  (Fig.~\ref{fig:spe}). These features have been identified as cyclotron
  resonant scattering features (CRSFs) by \citet{makishima92} and have also
  been reported by \citet{mihara95} and \citet{cusumano98}. The addition of a
  line at 36.0\,\kev\ with a width of 3.7\,\kev\
  (Table~\ref{tab:fitfdcut}) with a Gaussian optical depth profile \citep[for
  definition, see][]{coburn02} to the baseline continuum model results in a
  improved fit ($\chi_\text{red}^{2}= 2.49$ for 47 degrees of
  freedom). The $F$-test probability that this improvement is achieved just
  by chance is $3.8\times10^{-3}$ (see, however,
  \citealt{protassov02}).  Modeling both features with Gaussians at 39.8\,\kev\
  and 18.9\,\kev\ further improves $\chi^2$ significantly to
  $\chi^2_\text{red}=1.00$. The $F$-Test probability for this case as
  opposed to the case without lines is $9.8\times10^{-10}$, confirming
  the unambiguous detection of the cyclotron lines. For this model we
  also added the Fe-line which has been seen, e.g., in the \sax\
  data \citep{cusumano98}. The best fit value for the line energy is
  $7.1\,\text{keV}$. The width of the line could not be constrained by our
  data, so we fixed it to 0\,eV, corresponding to an equivalent width of
  135\,eV.

To exclude the possibility that the CRSFs are due to incorrect
modeling of the continuum, we also used other continuum
models\footnote{We did not use the \texttt{highecut} model due to the
  artificial features introduced in the fit residuals by the abrupt
  onset of the cutoff \citep{kretschmar97,kreykenbohm99}.} including
the NPEX model, which consists of negative and positive power laws
with a common exponential cutoff factor \citep{mihara95}:
\begin{equation}
\label{eq:npex}
  C(E)=A_{1}\left( E^{-\alpha_{1}}+A_{2}\cdot E^{+\alpha_{2}}\right)
  \,\exp \left(-\frac{E}{E_\text{fold}}\right)
\end{equation}
Using an absorbed NPEX model, we obtain $\chi^2_\text{red}=4.54$ and also
observe line like features at 19 and 40\,keV. Again we first fitted only one
feature at 44.4\,\kev, resulting in a $\redchi=2.56$.  Fitting these features
with two Gaussians at 42.0\,\kev\ and 18.7\,\kev\ improves the fit significantly
resulting in $\redchi=1.16$ ($F$-Test: $8.7\times10^{-13}$). See
Table~\ref{tab:fitnpex} for the complete set of model parameters. With the NPEX model it was not possible to constrain the Fe-line,
however the upper limit for a line with the same energy and sigma as in the
Fermi-Dirac cutoff model is
$9.7\,10^{-4}\,\text{photons}\,\text{cm}^{-2}\,\text{s}^{-1}$, consistent with
the results obtained for the continuum with a Fermi-Dirac cutoff.  The
continuum and line parameters measured with \inte\ are consistent with
the \sax\ results \citep{cusumano98}.  Our measurements also
confirm the \sax\ result that in this source the second line is
deeper than the fundamental line, a fact that can be explained by ``photon
spawning'', i.e., Compton scattering with multiple photon emission where a
photon with energy $2E_\text{cyc,1}$ results in two photons with energy
$E_\text{cyc,1}$ \citep{alexander89}. From our spectral fitting we
obtain an average flux of
$2.6\,10^{-10}\,\text{erg}\,\text{cm}^{-2}\,\text{s}^{-1}$ in the 20--40\,keV
band.

\begin{table}
  \caption{Best fit parameters for the Fermi-Dirac cutoff
    model. Shown are the 
    best-fit absorbing column, $N_\text{H}$, the photon index,
    $\Gamma$, the flux at 1\,keV, $A$, the cut-off energy,
    $E_\text{cut}$, the folding energy, $E_\text{fold}$, and the
    cyclotron line parameters (energy, $E_\text{cyc}$, depth
    $D_\text{cyc}$, and width, $\sigma_\text{cyc}$), and the reduced
    $\chi^2$ and the number of degrees of freedom, dof. All
    uncertainties are at the 90\% level for one interesting parameter.}
\label{tab:fitfdcut}
\centering
\begin{tabular}{llll}
\hline\hline
& \multicolumn{3}{c}{fdcut}\\
                                           & without line             &with 1
line & with 2 lines \\\hline\\
${N}_{\text{H}}$ [$10^{22}\,\text{cm}^{-2}$] & $3^{+7}_{-3} $           &$10^{+6}_{-7} $   & $0^{+7}_{-0}  $\\[1ex]
$\Gamma$                                   & $2.01^{+0.17}_{-0.10} $  &$2.21^{+0.15}_{-0.17} $  & $1.67^{+0.39}_{-0.09}   $\\[1ex]
$A$                                        & $0.21^{+0.13}_{-0.05}$ & $0.34^{+0.20}_{-0.13}$ &$0.11^{+0.03}_{-0.03} $\\[1ex]
$E_{\text{cut}}$ [keV]                     & $31^{+1}_{-1}$           &$37^{+6}_{-3}$        & $30^{+5}_{-8} $\\[1ex]
$E_{\text{fold}}$ [keV]                    & $5^{+1}_{-1} $     &$4^{+1}_{-1} $     & $7^{+3}_{-2}   $\\[1ex]
$E_{\text{cyc,1}}$ [keV]                   & --                       &--                       & $18.9^{+0.6}_{-0.7}  $\\[1ex]
$D_{\text{cyc,1}}$                         & --                       &--                       & $0.35^{+0.10}_{-0.09}   $\\[1ex]
$\sigma_{\text{cyc,1}}$ [keV]              & --                       &--                       & $3.0^{+1.0}_{-0.7}   $\\[1ex]
$E_{\text{cyc,2}}$ [keV]                   & --                       &$36.0^{+3.6}_{-2.4}  $& $39.8^{+4.0}_{-2.7}  $\\[1ex]
$D_{\text{cyc,2}}$                         & --                       & $0.7^{+0.6}_{-0.2}   $&$0.9^{+0.2}_{-0.4}   $\\[1ex]
$\sigma_{\text{cyc,2}}$ [keV]              & --                       &
$3.7^{+2.1}_{-2.2}   $&$9.0^{+2.7}_{-3.3}   $\\[1ex]
$E_{\text{Fe}}$ [keV]                   & --                       & --  & $7.1^{+0.3}_{-0.3}  $\\[1ex]
$\sigma_{\text{Fe}}$ [eV]              & --            & --  &$0$ (fixed)\\[1ex]
$A_{\text{Fe}}$ [$10^{-4}$]                        & --            & --  &$6^{+3}_{-3}   $\\[1ex]
$\chi^{2}/\text{dof}$              &  $154.91 / 50 $           &  $116.82 / 47 $           &$41.90 / 42 $\\[1ex]
$\chi_{\text{red}}^{2}$              &  $3.10 $           &  $2.49 $           &$1.00            $\\[1ex]
\hline
\end{tabular}
\end{table}

\begin{table}
\caption{Best fit parameters for the \texttt{npex} model, i.e., the
  photon index of the lower energy power law, $\alpha_1$ (the positive 
  power law index, $\alpha_2$, was fixed at 2.0), the power law normalizations $A_{1}$ and
  $A_{2}$ as defined in Eq.~\ref{eq:npex}, and the folding energy
   $E_\text{fold}$. All other symbols have the same meaning as in Table~\ref{tab:fitfdcut}.} 
\label{tab:fitnpex}
\centering
\begin{tabular}{llll}
\hline\hline
& \multicolumn{3}{c}{npex}\\
                                           & without line              & with 1 line          & with 2 lines \\\hline\\
${N}_{\text{H}}$ [$10^{22}\text{cm}^{-2}$] & $1^{+8}_{-1} $            & $4^{+9}_{-4} $            &$3^{+7}_{-3}   $\\[1ex]
$\alpha_{1}$                               & $0.88^{+0.20}_{-0.13} $ &$0.98^{+0.22}_{-0.15} $ & $0.87^{+0.29}_{-0.20}  $\\[1ex]
$A_{1}$ [$10^{-2}$]                        & $9^{+7}_{-1}$ &$9^{+7}_{-3}$ & $8^{+7}_{-3}$\\[1ex]
$A_{2}$ [$10^{-4}$]                        & $7^{+1}_{-3}$ &$2^{+1}_{-1}$ & $6^{+6}_{-3}$\\[1ex]
$E_\text{fold}$ [keV]                      & $4.85^{+0.04}_{-0.03} $   &$6.21^{+0.08}_{-0.10} $   & $5.16^{+0.76}_{-0.33}   $\\[1ex]
$E_{\text{cyc,1}}$ [keV]                   & --                        & --                        & $18.7^{+0.6}_{-0.8}  $\\[1ex]
$D_{\text{cyc,1}}$                         & --                        & --                        & $0.32^{+0.20}_{-0.08}   $\\[1ex]
$\sigma_{\text{cyc,1}}$ [keV]              & --                        & --                        & $2.8^{+1.1}_{-0.9}   $\\[1ex]
$E_{\text{cyc,2}}$ [keV]                   & --                        & $44.4^{+2.5}_{-2.0}  $&$42.0^{+5.1}_{-3.6}  $\\[1ex]
$D_{\text{cyc,2}}$                         & --                        & $1.32^{+0.37}_{-0.29}   $&$0.77^{+0.78}_{-0.37}   $\\[1ex]
$\sigma_{\text{cyc,2}}$ [keV]              & --                        & $7.0^{+1.5}_{-1.2}   $&$7.8^{+4.6}_{-2.7}   $\\[1ex]
$\chi^{2}/\text{dof}$              & $227.05 / 50  $            & $120.46 / 47  $            &$50.88 / 44            $\\[1ex]
$\chi_{\text{red}}^{2}$              & $4.54  $            & $2.56  $            &$1.16            $\\[1ex]
\hline
\end{tabular}
\end{table}

\section{Timing Analysis}
\subsection{The X-ray light curve of \fu}\label{sec:lc}
The X-ray light curve of \fu\ shows a clearly pulsed signal with a
period of $\sim$441\,s.  The mean count rates in the 20--40\,\kev\ band
for our observations are shown in Table~\ref{tab:obs}. The variations
in count rate largely reflect the orbital phase dependent variability
(Fig.~\ref{fig:asm}). 

\begin{figure}  
  \resizebox{\hsize}{!}{\includegraphics{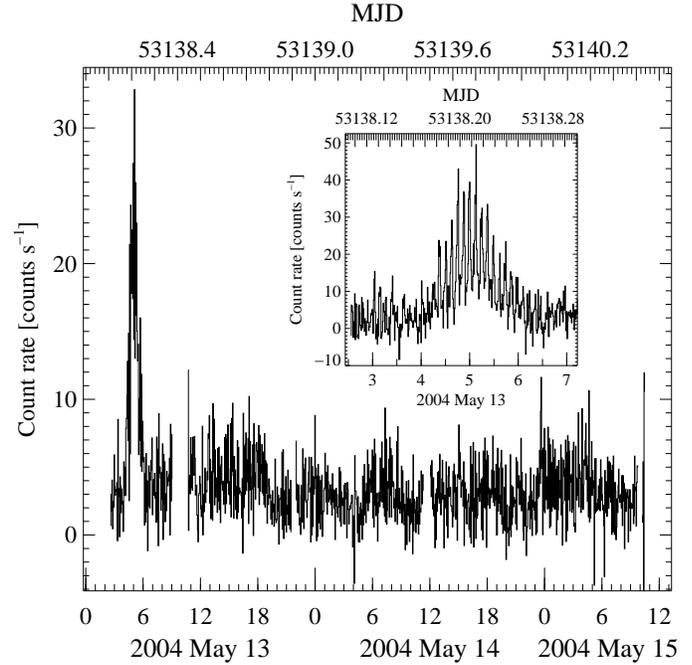}}
  \caption{IBIS (ISGRI) 20--40\,\kev\ light curve of revolution 193. The bin
  time is 200\,s. The inset shows a close up of the flare, the bin
  time in this case is 40\,s. X-axis numbers are hours.}
   \label{fig:193lcinset}
\end{figure}

\fu\ is also known to exhibit flares on a timescale of hours
\citep{makishima84,intzand98,mukerjee01}.  We observed four such
flares with count rate increases of at least $5\sigma$ over its normal
level (Fig.~\ref{fig:193lcinset}). Three of them, in revolutions 187,
193, and 305, are associated with the main peak in the orbital
lightcurve (Fig.~\ref{fig:asm}) while the flare observed in revolution
188 is linked to the secondary peak.  Table \ref{tab:flares} shows an
overview of the properties of these flares.

\begin{table}
\caption{Flares in the X-ray light curve of \fu. The peak rates are quoted for
a 441\,s binned light curve in the 20--40\,\kev\ band.}
\label{tab:flares}
\centering
\begin{tabular}{ccccc}
\hline\hline
Revolution & Center of Flare & Duration & $\Phi_\text{orb}$ & Peak
  Rates  \\
    & MJD      & sec & & IBIS $\text{cts}\,\text{s}^{-1}$\\
\hline
187 & 53121.65 & 7000  & 0.63 & 18.09\\ 
188 & 53125.03 & 5000  & 0.03 & 12.11\\ 
193 & 53138.21 & 7000  & 0.61 & 22.56\\ 
305 & 53472.93 & 2500  & 0.57 & 13.88\\
\hline
\end{tabular}
\end{table}

During flares, \citet{intzand98} and \citet{mukerjee01} also reported the
detection of transient 18.2\,s and 14.4\,s quasi-periodic oscillations
(QPOs). We therefore considered to calculate dynamical power spectra for the flare
light curves to search for similar oscillations, however, simulated light
curves showed that our \inte\ data are not sensitive enough to detect a QPO at
the reported levels.

\begin{figure}  
 \resizebox{\hsize}{!}{\includegraphics{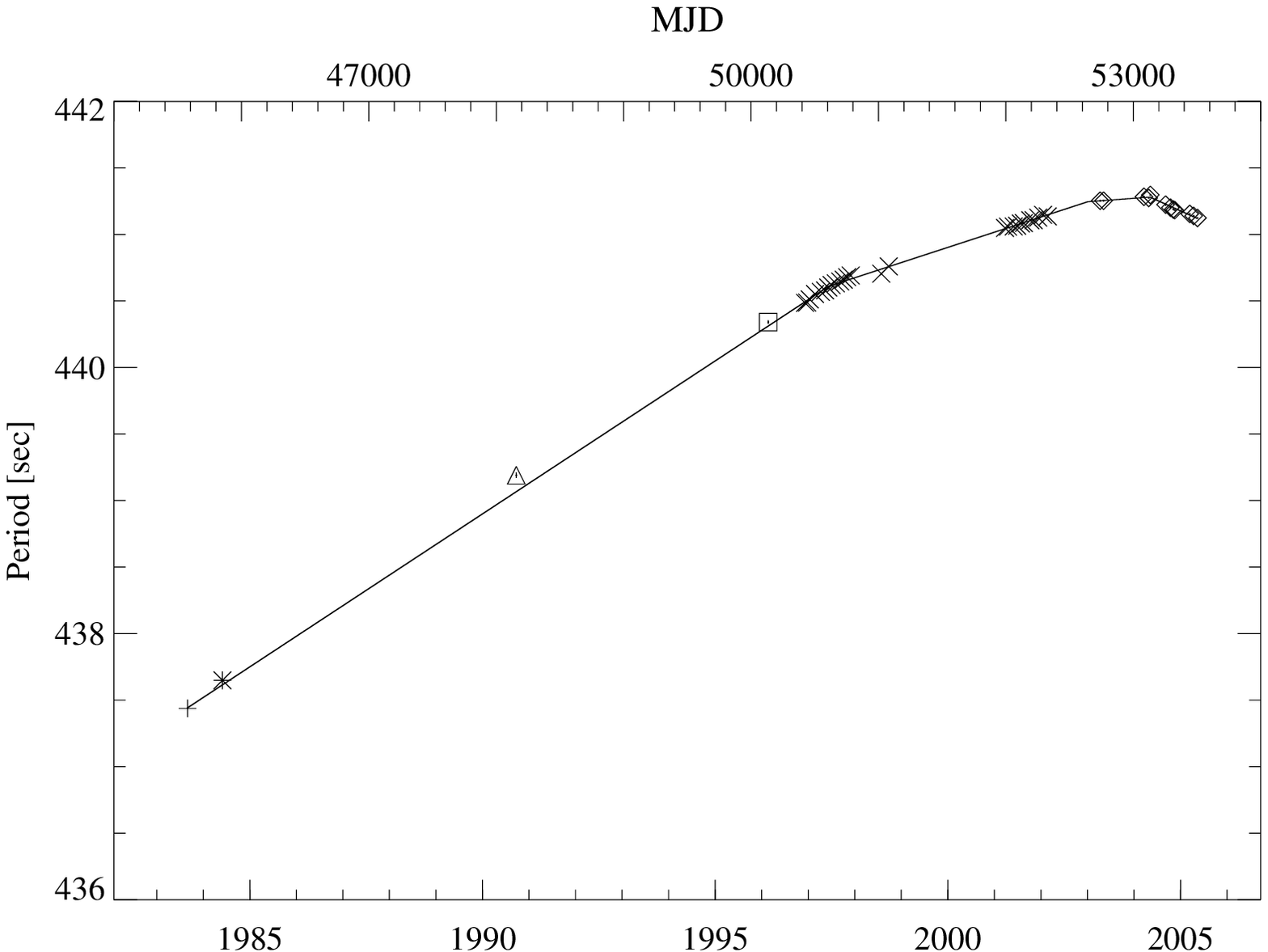}}
 \caption{Evolution of the \fu\ pulse period over the last 20 years. Diamonds
   indicate the results obtained in this work, other symbols
     indicate data from \citet[][plus-signs]{makishima84},
     \citet[][asterixes]{cook87}, \citet[][triangles]{mihara95}, \citet[][squares]{intzand98}, and
     \citet[][crosses]{baykal01,baykal05}. From MJD\,51980 onwards
   \fu\ shows a clear deviation from the long term spin down trend with
   a complete reversal to a spin up after MJD\,53131. The solid line
   shows the fit of four distinct episodes of different, but constant,
   $\dot{P}$. }
   \label{fig:perievo}
\end{figure}

\subsection{Pulsar Period and Pulse Profiles}\label{sec:pulsevol}
It is possible to track the evolution of the pulse period of \fu\ for
more than 20 years. Since the first measurements, the source showed a
steady spin down rate of
$\dot{P}_\text{pulse}=+0.225\,\text{s}\,\text{yr}^{-1}$
\citep{intzand98}.  A recent analysis of \xte\ data, however, taken
during MJD\,51980--52340 showed a spin down rate which is 0.5 times
lower than the previous value \citep{baykal05}. Using our \inte\ data,
we can extend the pulse period evolution to the years 2003--2005 (MJD
52739--53504).

\begin{figure}  
  \resizebox{\hsize}{!}{\includegraphics{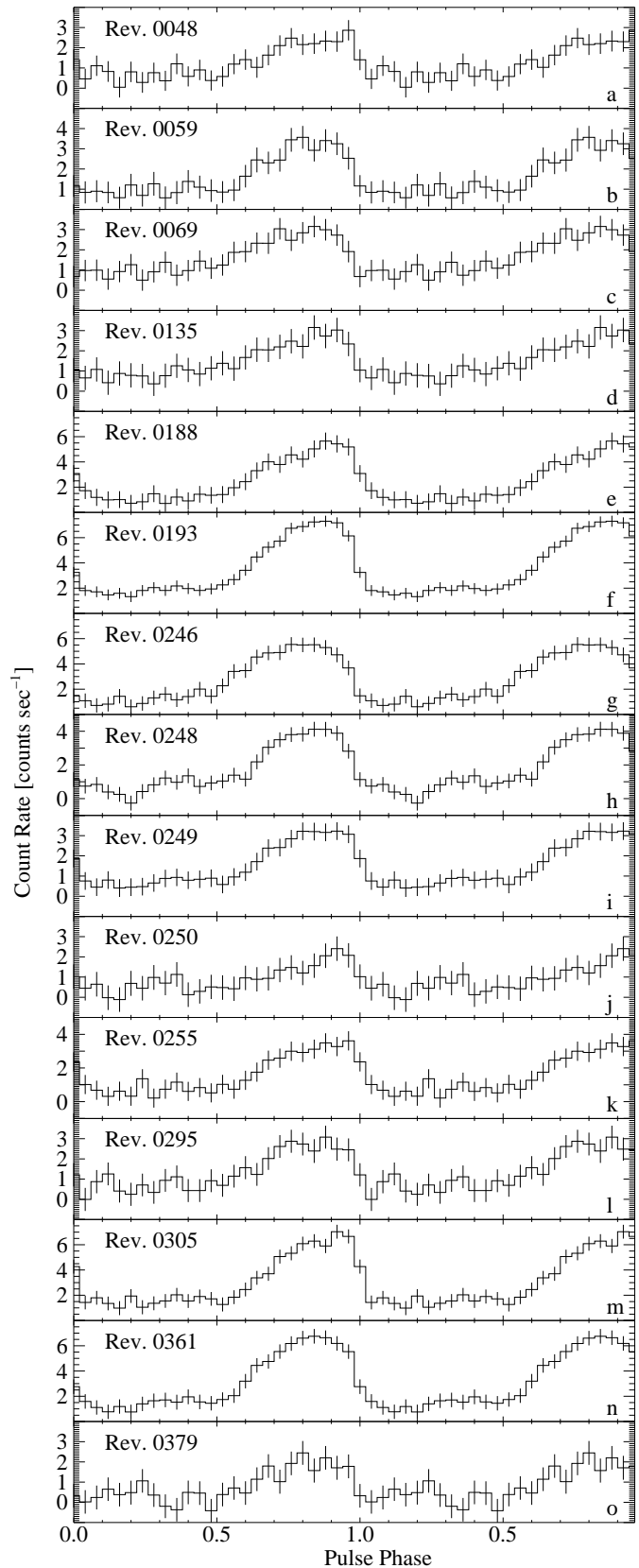}}
  \caption{Pulse profiles for different revolutions in the 20--40\,\kev\ band. Panels a--e show 
    pulse profiles obtained during the period of spin down, and panels f--o 
    show spin up pulse profiles. }
   \label{fig:pp_rev}
\end{figure}

To measure the pulse period of \fu\ we first performed epoch folding
\citep{leahy83} on each revolution separately. Using these periods we derived
pulse profiles for each revolution using the respective period. The pulse
profiles were then fitted by a sum of sine functions to determine the position
of the trailing edge (see for example Fig~\ref{fig:pp_rev} at phase 1.0) of
the pulse to obtain absolute pulse arrival times and the differences between
the arrival times of adjacent revolutions. From these time differences the
approximate number of pulse cycles between those absolute times was
determined. For sufficiently well sampled observations, those values are close
to an integer number. These integer numbers are then taken to count
the cycles throughout those well sampled subsets, and a model can be
constructed to calculate expected pulse arrival times (trailing edges). The
observed times can then be compared with the calculated times and a polynomial
fit to the observed times yields values for the period and its derivative.
Care was taken that within subsets no miscounting was possible.  Between
subsets the time gaps are too large such that they cannot be bridged uniquely
and no statement about the period development during the gaps can be made.
Using this procedure we were able to derive 12 periods with sufficient
accuracy (see Table~\ref{tab:periods}). 

\begin{figure}  
  \resizebox{\hsize}{!}{\includegraphics{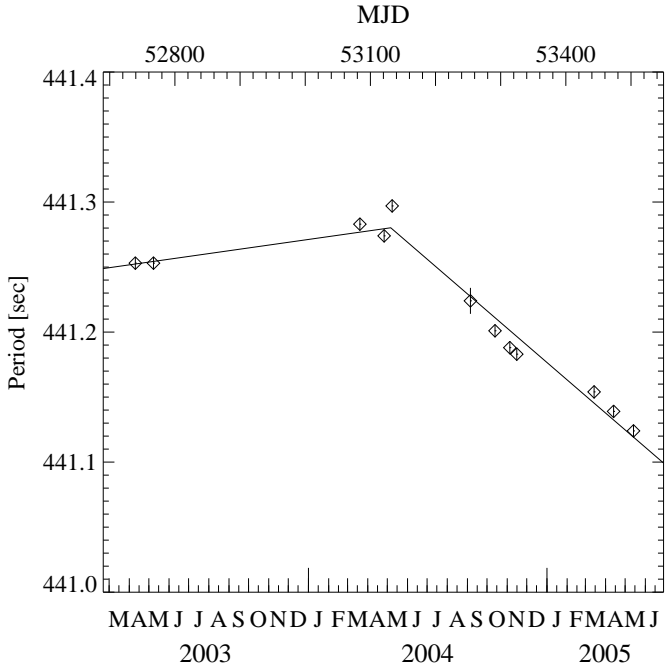}}
  \caption{Close up of the periods found in this work. Between MJD\,52643 to
     MJD\,53131 the derived pulse periods show a spin down of the source with
     $\dot P_\text{pulse}=+0.026\,\text{s}\,\text{yr}^{-1}$. After MJD\,53131
     the accretion torque changes sign and \fu\ spins up with $\dot
     P_\text{pulse} = -0.158\,\text{s}\,\text{yr}^{-1}$. The line indicates
     the linear fit to the periods as described in the text.}
   \label{fig:inteperievo}
\end{figure}

Fig.~\ref{fig:perievo} shows the long-time history of the period evolution
based on all available data. The line indicates the historic spin down
trend
with $\dot P_{\text{pulse}}=+0.225\,\text{s}\,\text{yr}^{-1}$ from
\citet{intzand97}. The recent periods obtained by \citet{baykal05} show a
clear deviation from the historic spin down trend. The periods obtained from
\inte\ (shown as diamonds), not only confirm this change of the trend but show
a complete trend reversal from the historic long term spin down trend to a
spin up trend from MJD\,53131 onwards (see Figs.~\ref{fig:perievo}
and~\ref{fig:inteperievo}). 

We tried several models to fit this overall period evolution\footnote{For this
analysis we excluded the two period measurements by \textsl{IXAE}
\citep{mukerjee01} as outliers.  Including them resulted in
$\chi_\text{red}^{2}$ values increased by a factor of 5.}. The description of
the data by three distinct episodes with a linear spin down from MJD\,45576 to
MJD\,50290 ($\dot{P}_\text{pulse}=+0.235\pm0.001\,\text{s}\,\text{yr}^{-1}$),
a parabolic turnover from MJD\,50290 to MJD\,53022 of the form
$P(T)=P_0+\dot{P}_\text{pulse,0} (T-T_0) + \ddot{P}_\text{pulse,0}
(T-T_0)^2/2$ with $T_0=\text{MJD}\,51321$,
$\dot{P}_\text{pulse,0}=+0.121\pm0.001\,\text{s}\,\text{yr}^{-1}$ and $\ddot{P}_\text{pulse,0}=-(7.743\pm0.002)\,10^{-3}\,\text{s}\,\text{yr}^{-2}$
, and a linear spin up with
$\dot{P}_\text{pulse}=-0.147\pm0.006\,\text{s}\,\text{yr}^{-1}$ from
MJD\,53022 onwards resulted in $\chi_\text{red}^{2}= 8.74$ for 34 degrees of
freedom. Modeling the period evolution of \fu\ with four distinct episodes of
different, but constant, $\dot{P}$ results in a slightly better
$\chi_\text{red}^{2}$ ($\chi_\text{red}^2=7.2$ for 34 dof, see
Figs.~\ref{fig:perievo} and~\ref{fig:inteperievo}). The four episodes are the
historic spin down with
$\dot{P}_\text{pulse}=+0.230\pm0.001\,\text{s}\,\text{yr}^{-1}$ (MJD\,45576 to
MJD\,50610), the turnover with
$\dot{P}_\text{pulse}=+0.114\pm0.001\,\text{s}\,\text{yr}^{-1}$ (MJD\,50610 to
MJD\,52643) and
$\dot{P}_\text{pulse}=+0.026\pm0.003\,\text{s}\,\text{yr}^{-1}$ (MJD\,52643 to
MJD\,53131), and finally the spin up with
$\dot{P}_\text{pulse}=-0.158\pm0.007\,\text{s}\,\text{yr}^{-1}$.

\begin{table}
  \caption{Period measurements of \fu. Historic measurements are taken from
    \citet[][M84]{makishima84}, \citet[][C87]{cook87},
    \citet[][M95]{mihara95}, \citet[I98]{intzand98},
    \citet[][M01]{mukerjee01}, \citet[][B01]{baykal01}, and
    \citet[][B05]{baykal05}. } 
\label{tab:periods}
\centering
\begin{tabular}{llll}
\hline\hline
Date     & Instrument     & Pulse Period       & Reference\\
MJD      &                & s                  & \\
\hline
 45576.5 & \textsl{Tenma} & $437.438 \pm0.004$ & M84\\
 45850.7 & \textsl{EXOSAT}& $437.649 \pm0.019$ & C87\\
 48156.6 & \ginga         & $439.19  \pm0.02$  & M95\\
 50134.8 & \xte           & $440.341 \pm0.014$ & I98\\
 50302.0 & \textsl{IXAE}  & $440.53  \pm0.01$  & M01\\
 50424.3 & \xte           & $440.4854\pm0.0109$& B05\\
 50440.4 & \xte           & $440.4877\pm0.0085$& B01\\
 50460.9 & \xte           & $440.5116\pm0.0075$& B05\\
 50502.1 & \xte           & $440.5518\pm0.0053$& B05\\
 50547.1 & \xte           & $440.5681\pm0.0064$& B05\\
 50581.1 & \xte           & $440.5794\pm0.0097$& B05\\
 50606.0 & \xte           & $440.6003\pm0.0115$& B05\\
 50631.9 & \xte           & $440.6189\pm0.0089$& B05\\
 50665\phantom{.}\phantom{0}  
         & \textsl{IXAE}  & $440.95  \pm0.01$  & M01\\
 50665.5 & \xte           & $440.6323\pm0.0069$& B05\\
 50699.4 & \xte           & $440.6460\pm0.0087$& B05\\
 50726.8 & \xte           & $440.6595\pm0.0105$& B05\\
 50754.1 & \xte           & $440.6785\pm0.0088$& B05\\
 50782.5 & \xte           & $440.6910\pm0.0097$& B05\\
 51021.9 & \xte           & $440.7045\pm0.0032$& B01\\
 51080.9 & \xte           & $440.7598\pm0.0010$& B01\\
 51993.8 & \xte           & $441.0484\pm0.0072$& B05\\
 52016.8 & \xte           & $441.0583\pm0.0071$& B05\\
 52061.5 & \xte           & $441.0595\pm0.0063$& B05\\
 52088.0 & \xte           & $441.0650\pm0.0063$& B05\\
 52117.4 & \xte           & $441.0821\pm0.0062$& B05\\
 52141.2 & \xte           & $441.0853\pm0.0082$& B05\\
 52191.4 & \xte           & $441.1067\pm0.0046$& B05\\
 52217.2 & \xte           & $441.1072\pm0.0077$& B05\\
 52254.3 & \xte           & $441.1259\pm0.0074$& B05\\
 52292.0 & \xte           & $441.1468\pm0.0065$& B05\\
 52328.8 & \xte           & $441.1353\pm0.0090$& B05\\
 52739.3 &\inte           & $441.253\pm0.005$  & this work\\
 52767.1 &\inte           & $441.253\pm0.005$  & this work\\
 53083.9 &\inte           & $441.283\pm0.005$  & this work\\
 53121.1 &\inte           & $441.274\pm0.005$  & this work\\
 53133.4 &\inte           & $441.297\pm0.005$  & this work\\
 53253.6 &\inte           & $441.224\pm0.010$  & this work\\
 53291.3 &\inte           & $441.201\pm0.005$  & this work\\
 53314.0 &\inte           & $441.188\pm0.005$  & this work\\
 53324.7 &\inte           & $441.183\pm0.005$  & this work\\
 53443.4 &\inte           & $441.154\pm0.005$  & this work\\
 53473.3 &\inte           & $441.139\pm0.005$  & this work\\
 53503.8 &\inte           & $441.124\pm0.005$  & this work\\
\hline
\end{tabular}
\end{table}

\begin{figure}  
  \resizebox{\hsize}{!}{\includegraphics{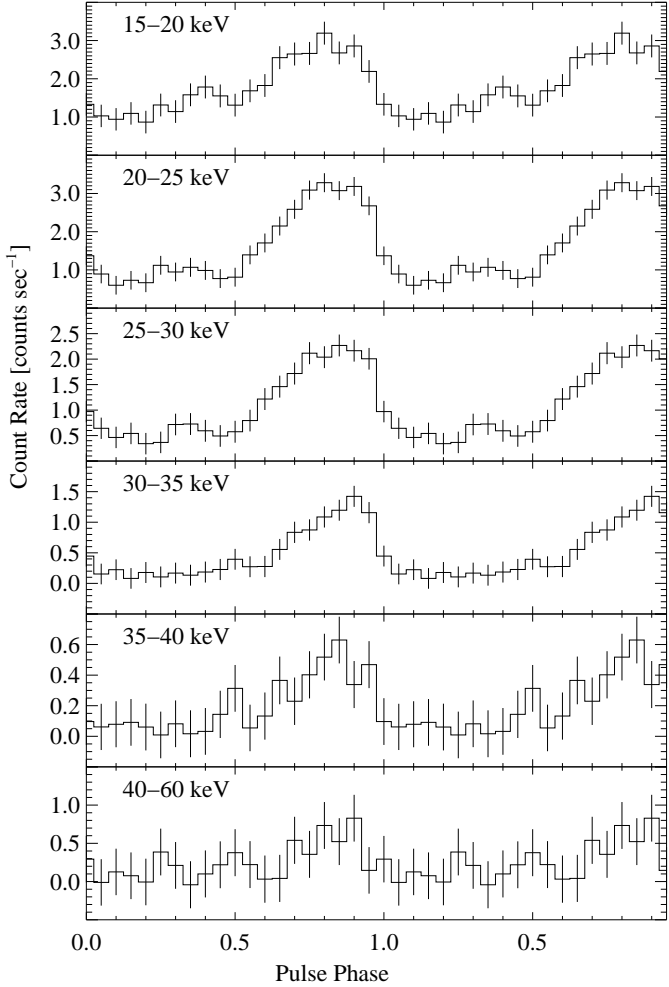}}
  \caption{Energy resolved pulse profiles for revolution 193
    ($P_\text{pulse}=441.224$\,s).}  
   \label{fig:pp}
\end{figure}

Using the derived periods, pulse profiles for all revolutions were
determined (Fig.~\ref{fig:pp_rev}) by folding the light curves with
the best respective periods. No change in the shape of the pulse
profile is detectable, and the \inte\ pulse profiles before and after the
reversal of $\dot{P}$ are consistent with each other. In addition, we
also obtained energy resolved pulse profiles (see Fig.~\ref{fig:pp}
for an example). Again, the evolution of the pulse profile from below
30\,keV to high energies is identical for all revolutions: the profile
exhibits a single peak at all energies with only a small secondary
hump, indicating the potential presence of a second peak at energies
$\lesssim$20\,keV, where a clear double peaked pulse profile was
observed previously \citep{makishima84,cook87,intzand98}.  At energies
above 60\,keV, the source gets too faint and no pulsed emission is
observable.

\section{Summary and Discussion}\label{sec:summary}
Since 2003, \fu\ was observed frequently with \inte. In this paper, we
presented the results of the timing analysis of $\sim$2280\,ksec of
IBIS data and of the spectral analysis of 1258\,ksec of IBIS and
200\,ksec of JEM-X2 data.  Two cyclotron lines at $\sim$19\,\kev\ and
$\sim$40\,\kev\ are detected, consistent with earlier \ginga\ 
and \sax\ results \citep{mihara95,cusumano98}.  Assuming a
gravitational redshift of $z=(1-(2GM/rc^{2}))^{-1/2}-1=0.31$ (assuming
$M=1.4\msun$ and $r=10^{6}\,\text{cm}$), from these CRSF observations
a $B$-field strength of $2.15\times10^{12}\,\text{G}$ can be inferred.
Four flares in the X-ray lightcurves with durations between
$\sim$2500\,s and $\sim$7000\,s were observed, three of which are
associated with the primary flare in the orbital lightcurve of \fu.

In addition to these confirmations of earlier results, our \inte\ data
show, for the first time, a clear spin up phase in \fu\
(Table~\ref{tab:periods} and in Figs.~\ref{fig:perievo}
and~\ref{fig:inteperievo}).  After almost 20 years of constant spin
down at a rate of
$\dot{P}_{\text{pulse}}=+0.225\,\text{s}\,\text{yr}^{-1}$, with the
available data being consistent with $\ddot{P}=0\,\text{s}^{-1}$, in
2002 and 2003 first indications for a decrease in the magnitude of
spin down were detected \citep{baykal05}.  As shown in
Sect.~\ref{sec:pulsevol}, the new \inte\ data show a torque reversal
from $\sim$MJD\,53131 onwards, with the source now exhibiting a spin
up with a rate of
$\dot{P}_\text{pulse}=-0.158\,\text{s}\,\text{yr}^{-1}$.  So far, the
\inte\ results appear to be consistent with
$\ddot{P}=0\,\text{s}^{-1}$ (the $1\sigma$ upper limit for $\ddot{P}$
  during the spin up is $-4\,10^{-5}\,\text{s}^{-1}\,\text{yr}^{-2}$).

The traditional interpretation of spin up in X-ray binaries with a
strongly magnetized, disk-accreting neutron star is that the accretion
disk is truncated at the inner edge of the disk by the magnetic field
of the neutron star \citep[see, e.g.,][and references
therein]{ghosh77,ghosh79a,ghosh79b}. For prograde disks, the magnetic
coupling between the accretion disk and the neutron star and the
transfer of angular momentum from the accreted matter onto the neutron
star will lead to a torque onto the neutron star resulting in spin up.
The general expectation of these models is that the accretion disk is
close to equilibrium, i.e., the Alfv\'en or magnetospheric radius,
$r_\text{m}$, where the accretion flow couples to the magnetic field,
is assumed to be close to the corotation radius, $r_\text{co}$, where
the disk's Kepler frequency equals the rotational frequency of the
neutron star. For accretion to occur, $r_\text{m}\lesssim
r_\text{co}$, as otherwise the onset of the centrifugal barrier (the
``propeller effect'') prevents the material from falling onto the
neutron star \citep{illarionov75}.

Most accreting neutron star systems, such as \object{Her~X-1} or
\object{Vela X-1}, show a secular spin up or spin down trend onto
which short, large magnitude episodes of spin down or spin up are
superimposed \citep{bildsten97}. These shorter episodes are often
interpreted to be due to short term variations in $\dot{M}$, which
give rise to short-term torque fluctuations. In contrast to these
systems, GX~1+4, 4U\,1626$-$67, and \fu\ showed decade-long phases
where $\dot{P}$ did not change its sign.  \object{GX~1+4} was seen to
spin up with a very short characteristic timescale, $P/\dot{P}\sim
40$\,years, from the early 1970s until the source flux dropped below
detectability in the \textsl{EXOSAT} era of the early 1980s. From 1985
onwards the source reemerged with a strong spin down
\citep{makishima88,mony91,chakrabarty97}. Recent \inte\ observations
show that this spin down has continued until at least 2004
\citep{ferrigno06}.  A second source with a long phase of unchanged
sign of $\dot{P}$ is the ultracompact low mass X-ray binary
\object{4U\,1626$-$67}, which was spinning up between 1977 and 1989,
followed by a torque reversal and an extended spin down phase starting
in 1991 \citep{chakrabarty97b}.  In contrast to GX~1+4, no luminosity
drop was seen during the torque reversal and with $|P/\dot{P}|\sim
5000$\,years the characteristic spin up and spin down timescales are
significantly larger.  Provided that the torque reversal of \fu\
described in this paper does not change back in the next years, in its
overall period behavior, \fu\ resembles 4U\,1626$-$67: both sources
have $|P/\dot{P}| \sim$ several thousands of years and no
change in luminosity was observed during the torque reversal. On the
other hand, however, \fu\ is a high mass X-ray binary with a much
longer orbital period, and thus shares similarities with the low mass
X-ray binary GX~1+4, and furthermore \fu\ switched from spin down to
spin up, while GX~1+4 and 4U~1626$-$67 switched from spin up to spin
down.  Common for the three sources is that the magnitude of
$|P/\dot{P}|$ is similar for the spin up and the spin down.

With the simple magnetic torquing model outlined above, the long
distinct episodes of roughly constant $\dot{P}$ and the magnitude and
sign of $\dot{P}$ are difficult to explain: As discussed above, for
systems close to equilibrium it is expected that $r_\text{m}\sim
r_\text{co}$. In reality, however, this is not the case.  In the case
of \fu, it was first pointed out by \citet{intzand98} that the
magnetospheric radius inferred from the cyclotron line measurements is
$r_\text{m}\sim 2400$\,km, while $r_\text{co}\sim12000$\,km. Moving
$r_\text{m}$ out to $r_\text{co}$ would require a magnetic field of
$\sim$$10^{14}\,\text{G}$, which is two orders of magnitude larger
than the magnetic field deduced from the observed cyclotron lines.
Alternatively, $\dot{M}$ could be significantly lower than the value
inferred from the X-ray luminosity of the system. Assuming the
efficiency of accretion is comparable to other systems, however, a low
$\dot{M}$ would imply a significantly smaller distance to \fu\ than
allowed by its optical reddening, which seems equally unlikely.
\citet{intzand98} argued that the presence of quasi-periodic
oscillations (QPOs) and the long term systematic trend in the spin
down are strong indicators for the presence of a 
accretion disk in the system with a small inner disk radius.  In analogy to
GX~1+4 \citep[][ and references therein]{chakrabarty97}, these authors show
that a transient retrograde disk with a duty cycle of 1--5\% could provide a
sufficiently strong torque to explain the observed spin down.  This torquing
would coincide with a short term increase of source
luminosity. \citet{intzand98} speculate that the X-ray flares indicate the
presence of the retrograde disk, since a switch to a retrograde disk would
imply an increase in X-ray luminosity. They point out, however, that the
$\sim$1000\,s duration of the flares seen in their \textsl{RXTE} data is too
small with respect to the required duty cycle for a retrograde disk.  We note
that the \inte\ results show flare durations of several ksec duration
(Tab.~\ref{tab:flares}), which is more in line with the expected duty
cycle. Furthermore, the folded lightcurve of \fu\ indicates a brightening of
the source around phase 0.6 (Fig.~\ref{fig:asm}). With $0.1P_\text{orb}$
duration, this flare is long enough to easily accommodate the required duty
cycle of the retrograde disk and to explain the observed $\dot{P}$. We note,
however, that not all of the brightening seen in Fig.~\ref{fig:asm} is due to
the flares identified by us as removing these flares from the data still
results in a clear peak at phase 0.6 in the folded lightcurve. Part of the
brightening could therefore be also due to an enhanced $\dot{M}$ during this
phase of the elliptical orbit of the neutron star, although we cannot firmly
exclude the presence of smaller flares which would not be picked up due to our
conservative definition of a flare.

While the model of a retrograde disk can explain the magnitude of
$P/\dot{P}$, it is more difficult to reconcile the torque reversal
found with \inte\ with this model. First of all, as discussed above,
the four intervals of different $\dot{P}$ represent distinct episodes
of different torques on the neutron star.  In the retrograde disk
model, it is difficult to understand why the characteristic spin up
and spin down timescales appear so similar. Furthermore, these three
phases would then correspond to different duty cycles of the
retrograde disk, resulting in a change in observed flux. Such a
behavior has been observed, e.g., in GX~1+4, where torque and
luminosity are correlated during the spin down phase
\citep{chakrabarty97}. No such correlations are observable in \fu,
where the flux has not appreciably changed during the \inte\ and the
pointed \xte\ observations \citep[see also][]{baykal05}. In addition,
an analysis of the orbit-averaged 2--12\,keV \xte\ \asm\ lightcurves
also does not reveal a change in the soft X-ray behavior of the
source, and neither is a change in the X-ray light curve with orbital
phase observed.

The \inte\ observations also rule out the application of the torque
reversals put forward by \citet{murray99} to explain GX~1+4. In this
model, the accretion disk is assumed to consist of several rings with
opposite rotation.  Between these rings a gap is created such that a
torque reversal should be accompanied by a minimum in the accretion
luminosity.  The model predicts a change in pulse profile from
``leading-edge bright'' during the spin down to ``trailing-edge
bright'' during the spin up. Both, a minimum in luminosity and a
change in the shape of the pulse profile are observed in GX 1+4
\citep{greenhill99}, but neither effect is present in \fu\ (see
discussion above and Fig.~\ref{fig:pp_rev}). 

Recently, \citet{perna06} presented a new \textit{Ansatz} to explain
torque changes in accreting neutron stars, which does not require the
presence of retrograde disks. This model makes use of the fact that
X-ray pulsars are oblique rotators, i.e., the neutron star's magnetic
field is tilted by an angle $\chi$ with respect to the axis of
rotation of the neutron star. Assuming the accretion disk is situated
in the neutron star's equatorial plane, for a ring of matter in the
accretion disk, the magnetic field strength then depends on the
azimuthal angle and thus the boundary of the magnetosphere is
asymmetric.  As shown by \citet{perna06}, such a configuration can
lead to regions in the disk where the propeller effect is locally at
work, while accretion from other regions is not inhibited (i.e., on a
ring on the disk, there are some regions with $r_\text{m}<r_\text{co}$
and other regions with $r_\text{m}>r_\text{co}$). This results in a
nonlinear dependence of the accretion luminosity from the $\dot{M}$
through the outer parts of the disk.  A nontrivial consequence of this
more realistic accretion geometry is that for values of $\chi$ greater
than a critical value, $\chi_\text{crit}$, limit cycles are present,
where cyclic torque reversal episodes are possible without a change in
$\dot{M}$. For typical parameters. $\chi_\text{crit}$ is between
$\sim$$25^\circ$ and $\sim$$45^\circ$.  We stress that the existence
of the limit cycles depends only on $\chi$ and on the pulsar's polar
magnetic field, $B$, and no variation of external parameters is
required to trigger torque reversals.  This fact is a big advantage of
this model over the models discussed above.

\citet{perna06} show that their model can explain the principal observed
properties of GX~1+4, including the luminosity drop during torque
reversal, except for the observed correlation between torque and
luminosity during the spin down phase. For 4U\,1626$-$67, assuming
$B=2.5\times 10^{12}$\,G and $\chi=68^\circ$, all observed properties
of the spin history of 4U\,1626$-$67 including the large and values of
$|P/\dot{P}|$ before and after the torque reversal and the virtually
unchanged luminosity of the source can be explained (in the model, the
required change in luminosity is only $\sim$5\%). For these model
parameters, the limit cycle of 4U\,1626$-$67 is predicted to have a
long period.

The similarity between the pulse histories of \fu\ and 4U\,1626$-$67
suggests that a similar model would also work for the former source.
We stress, however, that further quantitative work, which is outside
the scope of this paper, is clearly required. Specifically, from a
theoretical point of view we note that the model of \citet{perna06}
does not yet take changes in $\dot{M}$ due to the orbital eccentricity
into account and that the current version of the model assumes the
disk to be flat and in the equatorial plane of the neutron star, while
more realistic disks onto magnetized objects can be expected to be
warped and precessing \citep[e.g.,][and therein]{pfeiffer04}. From an
observational point of view, the donor stars in both systems are very
different, with 4U\,1626$-$67 being a low mass and \fu\ being a high
mass system.  Finally, we also emphasize that the torque reversal in
4U\,1626$-$67 was from a spin up to a spin down, while for \fu\ it was
from a spin down to a spin up. Since the torque reversal episodes are
cyclic, however, we do not expect this latter difference to be of
major importance.

In conclusion, the torque reversal of \fu\ with no associated drop in
luminosity and no change in the shape of the pulse profile seem to be
difficult to reconcile with models explaining the large $P/\dot{P}$
through the presence of a retrograde disk. On the other hand, the
proposal of \citet{perna06} to explain torque reversals through
accretion onto an oblique rotator seems to be able to explain the long
term constancy of $\dot{P}$ trends in \fu\ and 4U\,1626$-$67, the large
magnitude of $|P/\dot{P}|$, and the constancy of X-ray flux over the
torque reversal, although further and more detailed theoretical
studies are required. As a prediction of the model for systems such as
\fu\ and 4U\,1626$-$67 is that torque reversal episodes are rare
events, further monitoring of the pulse period of \fu\ is required to
determine whether this change is a sign of a long term torque reversal
or whether we are just observing a short deviation from a long term
spin down trend.

\begin{acknowledgements}
  We acknowledge the support of the Deutsches Zentrum f\"ur Luft- und
  Raumfahrt under grant numbers 50OR0302, 50OG9601, and 50OG0501, and
  by National Aeronautics and Space Administration grant
  INTEG04-0000-0010. This work is based on observations with
  \textsl{INTEGRAL}, an European Space Agency (ESA) project with
  instruments and science data centre funded by ESA member states
  (especially the PI countries: Denmark, France, Germany, Italy,
  Switzerland, Spain), Czech Republic and Poland, and with the
  participation of Russia and the USA.
\end{acknowledgements}

\bibliographystyle{aa}
\bibliography{mnemonic,aa_abbrv,diplom}

\end{document}